\documentclass[reprint,showpacs,amssymb,amsmath,aps,pra]{revtex4-1}
\usepackage{graphicx}
\usepackage{bm}
\usepackage{amsthm}
\usepackage{lipsum}

\begin{document}
\title{Factorization in spin systems under general fields and
separable ground state engineering}
\author{M. Cerezo, R. Rossignoli, N. Canosa}
\affiliation{Instituto de F\'{\i}sica de La Plata and Departamento de F\'{\i}sica,
Universidad Nacional de La Plata, C.C.67, La Plata (1900), Argentina}
\begin{abstract}

We discuss ground state factorization schemes in spin $S$ arrays with general
quadratic couplings under general magnetic fields, not necessarily uniform or
transverse. It is shown that given arbitrary spin alignment directions at
each site, nonzero $XYZ$ couplings between any pair and fields at each site always
exist such that the ensuing Hamiltonian has an exactly separable eigenstate
with the spins pointing along the specified directions. Furthermore, by
suitable tuning of the fields this  eigenstate can always be cooled down to a
nondegenerate ground state. It is also shown that  in open one-dimensional
systems with fixed arbitrary first neighbor couplings, at least one separable
eigenstate compatible with an arbitrarily chosen spin direction at one site
is always feasible if the fields at each site can be tuned. We demonstrate as
well that in the vicinity of factorization, i.e., for small perturbations in the
fields or couplings, pairwise entanglement reaches full range. Some
noticeable examples of factorized eigenstates are
unveiled. The present results open the way for separable ground state
engineering. A notation to quantify the complexity of a given type of solution
according to the required control on the system couplings and fields is introduced.
 \end{abstract}
\pacs{75.10.Jm, 03.67.Mn, 03.65.Ud, 64.70.Tg}
\maketitle

\section{Introduction}

Over the last two decades quantum information and quantum computation sciences
have experienced an extraordinary theoretical and experimental progress
\cite{NC.00,HR.06,Ve.06}. In particular, the possibility of performing quantum
information processing and quantum simulation tasks in  archetypal many-body
systems such as spin arrays has attracted considerable attention
\cite{RF.85,ALS.02,LD.08,AV.08,GC.97,BK.98,CZ.95,SS.97}. Furthermore, the
recent noticeable advances in control techniques of quantum systems have made
it possible  to engineer and simulate spin interactions by means of cold atoms
in optical lattices \cite{SB.11,LS.12,GA.14}, superconducting Josephson
junctions \cite{DS.13,ZI.14,vP.07,B.16} or trapped ions
\cite{GA.14,PC.04,KC.09,BR.12,KK.12,BS.12,SR.15}, leading  to an ever
increasing interest in these strongly correlated systems.

In this framework, it is well known that the exact eigenstates of interacting
spin systems in the presence of an external magnetic field are typically
entangled states. However, one of the most exceptional features of these
systems is that under certain conditions they can posses a completely separable
ground state (GS), i.e., a {\it factorized} GS which can be expressed as the
product of single spin states \cite{KU.82}. The remarkable phenomenon of
factorization has been thoroughly studied in spin systems immersed in a uniform
transverse field
\cite{Mu.85,T.04,Am.06,GI.07,GI.08,RCM.08,CRM.10,G.09,ARL.12,SC.13}, and
in finite anisotropic $XYZ$ spin chains the transverse factorizing field
has been shown \cite{RCM.08,CRM.10} to correspond to a GS $S_z$ parity
transition, the ensuing separable GS being two-fold degenerate. Recently, in
Ref.\ \cite{MRC.15} we studied $XYZ$ models with uniform nontransverse fields,
where it was  shown that a uniform, maximally aligned, nondegenerate,
completely separable GS can exist in both ferromagnetic and
antiferromagnetic-type systems for fields parallel to a principal plane of the
coupling.

In this work we show that if some control over the couplings or the fields is
feasible, then completely separable exact ground states can be engineered in
general $XYZ$-type systems. This point is important  for the first basic step
in most quantum information processes \cite{VC.05} and quantum simulation
schemes \cite{GA.14,SR.15}, since they are based on an initial fully separable
state of the qubits, assumed to be reached with high fidelity. Whenever such
initial state is only approximately achieved or is prone to decoherence,
additional error correction must be implemented \cite{CC.05}. Therefore, the
possibility of having an exactly separable GS at {\it finite} magnetic fields
{\it even in the presence of strong interactions between the spins}, is highly
desirable, specially if such GS is nondegenerate and can be well separated from
the remaining spectrum. Such possibility can be also useful in schemes for 
quantum annealing \cite{DC.08,MJ.11}. 

We first show that for arbitrary alignment directions at each site, compatible
nonzero $XYZ$ couplings between any pair of spins and concomitant {\it finite}
factorizing fields always exist such that the separable state is an exact
eigenstate of the system. Moreover, such state can always be made a
nondegenerate well separated GS by appropriately tuning the fields. In
addition, it is shown, remarkably, that for an arbitrary quadratic coupling
between two spins and an  arbitrarily chosen spin alignment direction of one of
the spins, there is always an alignment direction of the remaining spin
compatible with an exactly separable eigenstate. This result enables to
engineer separable GS in systems with arbitrary first neighbor couplings at
least in one-dimensional-type geometries, if fields can be tuned. Furthermore,
the factorizing fields for a single pair can be always chosen as {\it uniform},
though in general nontransverse. A complexity classification scheme for the
control required on the couplings and fields is accordingly introduced. This
general framework also allows to identify and prove the existence of nontrivial
separable eigenstates for certain couplings, fields and geometries, like the
spin-spiral-type solution which will be discussed. We also suggest two
experimental implementations for which the proposed methods could  be realized.
A final but not less important aspect is that the present general factorization
points, arising for  not-necessarily uniform couplings and non-transverse
fields, can also be associated to an entanglement transition: pairwise
entanglement, though obviously vanishing at factorization, will be shown to
reach {\it full range} in its vicinity if either the  fields {\it or couplings}
are perturbed, in agreement with previous results for uniform fields
\cite{Am.06,RCM.08,MRC.15}.

The general rigorous results are presented and demonstrated in Sec.\ \ref{2}
and the Appendix. Special examples of factorized eigenstates are discussed in
Sec.\ \ref{3}. The ensuing GS engineering schemes, complexity classification,
and experimental implementation are  discussed in \ref{4}. Conclusions are
finally drawn in \ref{5}.

\section{Exactly separable eigenstates\label{2}}
\subsection{Separability conditions for general quadratic couplings}

We consider an array of $N$ spins $S_i$, not necessarily equal, interacting
through general quadratic  couplings of arbitrary range in the
presence of a general magnetic field $\bm{h}^i=(h^i_x,h^i_y,h^i_z)$. The
Hamiltonian is
\begin{eqnarray}
H&=&-\sum_i \bm{h}^i\cdot\bm{S}_i-{\textstyle\frac{1}{2}}
 \sum_{i\neq j} \bm{S}_i\cdot {\cal J}^{ij}\bm{S}_j\label{H0}\\
&=&-\sum_{i,\mu}h^i_\mu S^\mu_i - {\textstyle\frac{1}{2}} \sum_{i\neq j,\mu,\nu}
 J_{\mu\nu}^{ij} S^\mu_i S^\nu_j\,, \label{H}\end{eqnarray}
where $i,j$ label the sites in the array, $S_i^\mu$, $\mu=x,y,z$, denote the
spin components at site $i$ and $J_{\mu\nu}^{ij}=J^{ji}_{\nu\mu}$ the coupling
strengths between spins at sites $i$ and $j$, with ${\cal J}^{ij}$ a matrix of
elements $J^{ij}_{\mu\nu}$. The $XYZ$ case corresponds to ${\cal J}^{ij}$
diagonal $\forall$ $i,j$ ($J^{ij}_{\mu\nu}=\delta_{\mu\nu}J^{ij}_\mu$). The
Hamiltonian (\ref{H0}) will  possess a {\it completely separable eigenstate} of
the form
\begin{equation}
|\Theta\rangle=\otimes_{i=1}^n R_i|0_i\rangle=
|\nearrow\rightarrow\swarrow\nwarrow...\rangle,\;\;
R_i=e^{-\imath \phi_i S^z_i}e^{-\imath \theta_i S^y_i}\,,
\label{Theta}
\end{equation}
where $|0_i\rangle$ is the local state with maximum spin along the $z$
axis ($S^z_i|0_i\rangle=S_i|0_i\rangle$) and $R_i$ is a rotation such that
the resulting spin alignment direction is
$\bm{n}_{i}=(\sin\theta_i\cos\phi_i,\sin\theta_i\sin\phi_i,\cos\theta_i)$, if
two sets of conditions are met \cite{MRC.15}. The first ones are the pairwise field
independent equations which relate the alignment directions with the exchange couplings:
\begin{eqnarray}
\bm{n}_i^{x'}\cdot{\cal J}^{ij}\bm{n}_j^{x'}&=&
\bm{n}_i^{y'}\cdot{\cal J}^{ij}\bm{n}_j^{y'}\,,
\;\;\;\;\bm{n}_i^{x'}\cdot{\cal J}^{ij}\bm{n}_j^{y'}=
-\bm{n}_i^{y'}\cdot{\cal J}^{ij}\bm{n}_j^{x'}\,.\label{1v0}
\end{eqnarray}
Here
$\bm{n}_{i}^{x'}=(\cos\theta_i\cos\phi_i,\cos\theta_i\sin\phi_i,-\sin\theta_i)$,
$\bm{n}_{i}^{y'}=(-\sin\phi_i,\cos\phi_i,0)$,  are the corresponding rotated
vectors orthogonal to $\bm{n}_i^{z'}=\bm{n}_i$, such that
$(\bm{n}_{i}^{x'},\bm{n}_{i}^{y'},\bm{n}_{i})$ forms an orthonormal triad.
Eqs.\ (\ref{1v0}) mean that the strengths
$J^{ij}_{\mu'\nu'}=\bm{n}^{\mu'}_i\cdot{\cal J}^{ij}\bm{n}^{\nu'}_j$ associated
with the rotated spin operators $S_i^{\mu'}=R_i S_i^\mu R_i^\dagger$
($\bm{S}_i\cdot{\cal J}^{ij}\bm{S}_j=\sum_{\mu,\nu}J^{ij}_{\mu'\nu'}S_i^{\mu'}
S_j^{\nu'}$) satisfy ${J}^{ij}_{x'x'}=J^{ij}_{y'y'}$ and
$J^{ij}_{x'y'}=-J^{ij}_{y'x'}$, ensuring that $H$ does not connect
$|\Theta\rangle$ with two spin excitations.

The second set are the local field dependent equations which determine the
factorizing fields $\bm{h}^i$ at each site:
\begin{equation}
\bm{h}^i=\bm{h}^i_{\parallel}+\bm{h}^i_{\perp}\,,
\label{ht}
\end{equation}
where $\bm{h}^i_{\parallel}=h^i_{\parallel}\bm{n}_i$ is  an {\it arbitrary} field
{\it  parallel} to the local spin alignment direction $\bm{n}_i$ and
\begin{eqnarray}
\bm{h}^i_{\perp}&=&\bm{n}_i\times\left(\bm{n}_i\times(\sum_j S_j
  {\cal J}^{ij}\bm{n}_j)\right) \label{hp}\,,\end{eqnarray}
is a field  {\it orthogonal} to the local alignment direction $\bm{n}_i$, which
represents the nontransverse factorizing field of {\it lowest} magnitude and
ensures that $H$ will not connect $|\Theta\rangle$ with single spin
excitations: Eq.\ (\ref{hp}) (equivalent to $\bm{h}^i_{\perp}=\sum_j
\bm{h}^{ij}_{\perp}$, with $\bm{h}^{ij}_{\perp}=-S_j[{\cal J}^{ij}\bm{n}_j
-\bm{n}_i(\bm{n}_i\cdot{\cal J}^{ij}\bm{n}_j)]$) implies $h^i_{\mu'}=-\sum_j
S_j J^{ij}_{\mu'z'}$ for $\mu'=x',y'$.

If Eqs.\ (\ref{1v0})--(\ref{hp}) are fulfilled, then
$H|\Theta\rangle=E_\Theta|\Theta\rangle$, with
\begin{eqnarray}
E_{\Theta}&=& -\sum_i\langle \bm{S}_i\rangle\cdot\bm{h^i}
-{\textstyle\frac{1}{2}}\sum_{i\neq j}
\langle \bm{S}_i\rangle\cdot{\cal J}^{ij}\langle\bm{S}_j\rangle\label{EE}\nonumber\\
&=&-\sum_i S_i h^i_{\parallel}-{\textstyle\frac{1}{2}}\sum_{j\neq i,\mu,\nu}
S_i S_j J^{ij}_{\mu\nu} n_{i\mu}n_{j\nu}\,,
\label{EE2}
\end{eqnarray}
where $\langle\bm{S}_i\rangle\equiv \langle
\Theta|\bm{S}_i|\Theta\rangle=S_i\bm{n}_i$. This energy is split into two
contributions: the first  one is associated with the parallel field components
$\bm{h}^i_{\parallel}=h^i_{\parallel}\bm{n}_i$ and the second one, independent
of $\bm{h}^i_{\parallel}$, with the couplings. The parallel components
$\bm{h}^i_{\parallel}$ can then be used to shift the energy of the factorized
state and hence, {\it to cool it down to a GS}, as discussed below.

\subsection{Fundamental properties \label{lem}}

We now provide five fundamental properties of the previous maximally aligned
separable eigenstates. For the sake of clarity proof details are presented in
the Appendix.

{\bf Lemma 1.} {\it If Eqs.\ (\ref{1v0})--(\ref{hp}) are satisfied , the state
$|\Theta\rangle$ given by Eq.\ (\ref{Theta}) will always become a nondegenerate
GS of $H$ for sufficiently strong yet finite  parallel fields
$\bm{h}^i_{\parallel}=h^i_{\parallel}\bm{n}_i$.}

{\it Proof:}  This result is apparent as no state $|\Psi\rangle$ orthogonal to 
$|\Theta\rangle$ will have an energy $\langle H\rangle_{\Psi}\equiv \langle
\Psi|H|\Psi\rangle$ which decreases more rapidly with the applied fields
$h^i_{\parallel}$ than $E_{\Theta}$, since
$\langle\Theta|\bm{S}_i\cdot\bm{n}_i|\Theta\rangle=S_i$ is maximum. Hence, a
finite threshold value $h_{\parallel c}$ {\it will always exist} such that
$|\Theta\rangle$ becomes a nondegenerate GS  if $h^i_{\parallel}>h_{\parallel
c}$ $\forall$ $i$. Moreover, the energy gap with the first excited state can be
made as large as desired by increasing the values of ${h}^i_{\parallel}$.\qed

{\bf Lemma 2.} {\it Given two arbitrary alignment directions $\bm{n}_i$,
$\bm{n}_j$ at sites $i,j$, a non-zero  $XYZ$-type   coupling
$J^{ij}_{\mu\nu}=J_\mu^{ij}\delta_{\mu\nu}$
 always exists such that Eqs.\ (\ref{1v0}) are fulfilled}.

This Lemma implies that for {\it arbitrary} alignment directions $\bm{n}_i$ at
each site of the array, $XYZ$ couplings $J_\mu^{ij}$  and suitable
fields $\bm{h}^i$ always exist
such that the associated factorized state $|\Theta\rangle$ is an exact GS of $H$.

\begin{figure}[t]
 \centerline{\scalebox{.9}{\includegraphics{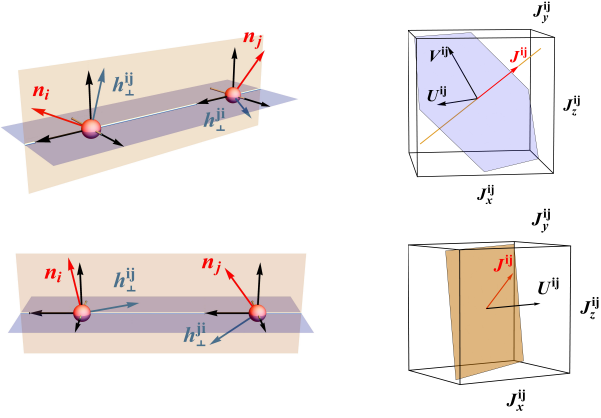}}}
\caption{(Color online). Left column: Schematic representation of the spin
alignment directions $\bm{n}_i$ and $\bm{n}_j$ and the perpendicular
factorizing fields $\bm{h}^{ij}_{\perp}$ and $\bm{h}^{ji}_{\perp}$ (Eq.\
(\ref{hp})). Right column: Exchange couplings in the exchange couplings
space compatible with the spin alignment configuration. Top row: $\bm{U}^{ij}$
and $\bm{V}^{ij}$ are linearly independent and Eqs.\ (\ref{2v})
define a {\it line} (Eq.\ (\ref{jc})) of possible exchange couplings. Bottom
row: $\bm{U}^{ij}$ and $\bm{V}^{ij}$ are linearly dependent (as $\bm{n}_i$ and
$\bm{n}_j$ belong to a principal plane) and Eqs.\ (\ref{2v}) define
a plane of possible exchange couplings (Eq.\ (\ref{jp})). }
 \label{f1}
\end{figure}

{\it Proof:} For $J^{ij}_{\mu\nu}=J^{ij}_{\mu}\delta_{\mu\nu}$, Eqs.\
(\ref{1v0}) can be conveniently rewritten in vector form as
\begin{eqnarray}
\bm{J}^{ij}\cdot\bm{U}^{ij}=0\,, \; \bm{J}^{ij}\cdot\bm{V}^{ij}=0\,,\label{2v}
\end{eqnarray}
where $\bm{J}^{ij}=(J^{ij}_x,J^{ij}_y,J^{ij}_z)$ is the exchange coupling
vector and $\bm{U}^{ij}$, $\bm{V}^{ij}$ the vectors defined as 
$\bm{U}^{ij}=\bm{n}^{x'}_i\ast\bm{n}^{x'}_j-\bm{n}^{y'}_i\ast\bm{n}^{y'}_j$, 
$\bm{V}^{ij}=\bm{n}^{x'}_i\ast\bm{n}^{y'}_j+\bm{n}^{y'}_i\ast\bm{n}^{x'}_j$,
with $\bm{n}\ast\bm{m}=(n_xm_x,n_y m_y,n_z m_z)$ the Hadamard product, such that
$\bm{n}_i\cdot{\cal J}^{ij}\bm{n}_j=\bm{J}\cdot(\bm{n}_i\ast\bm{n}_j)$ for ${\cal J}^{ij}$ diagonal.  
Hence, by choosing ${\bm J}^{ij}$ {\it orthogonal} to the subspace generated by
$\bm{U}^{ij}$ and $\bm{V}^{ij}$, Eqs.\  (\ref{2v}) (and then (\ref{1v0})) are
fulfilled. The fields can then be obtained from Eqs.\ (\ref{ht})--(\ref{hp}).
And by applying sufficiently strong parallel fields $\bm{h}^i_{\parallel}$,
$|\Theta\rangle$ can be made a GS (Lemma 1).  \qed

Note that two distinct situations are implied by Eq.\ (\ref{2v}), as depicted
in Fig.\ \ref{f1}: If $\bm{U}^{ij}$ and $\bm{V}^{ij}$ are {\it linearly
independent}, it determines a {\it line} of compatible exchange vectors
orthogonal to the plane generated by $\bm{U}^{ij}$ and $\bm{V}^{ij}$, i.e.,
 \begin{equation}
\bm{J}^{ij}=j^{ij} (\bm{U}^{ij}\times\bm{V}^{ij})\,,
\label{jc}
\end{equation}
 with $j^{ij}$ an arbitrary real constant.

On the other hand, if $\bm{U}^{ij}$ and $\bm{V}^{ij}$ are {\it linearly
dependent}, it defines a {\it plane} of compatible exchange couplings.  This
case arises  whenever i) $\bm{n}_i$ and $\bm{n}_j$ belong to the same principal
plane (i.e., $n_{j\sigma}=n_{i\sigma}=0$ for some $\sigma=z$, $x$ or $y$), ii)
$\bm{n}_j$ is the reflection of $\bm{n}_i$ with respect to a principal plane
($|n_{j\mu}|=|n_{i\mu}|$ $\forall$ $\mu$, with  $n_{j\sigma}=-n_{i\sigma}$ for
just one component $\sigma$) and iii) $\bm{n}_{i}=-\bm{n}_{j}$, i.e.,
antiparallel alignment directions. In these cases $\bm{V}^{ij}$ vanishes for
the present choice of orthogonal vectors $\bm{n}_i^{\mu'}$ and the plane of
compatible exchange couplings is that orthogonal to $\bm{U}^{ij}$. The explicit
expressions for $\bm{J}^{ij}$ are given in the Appendix.

\begin{figure}[t]
\centerline{\scalebox{.9}{\includegraphics{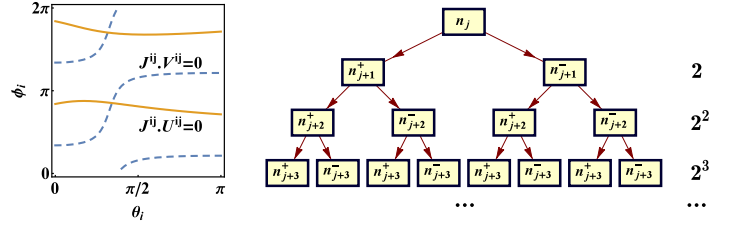}}}
 \caption{(Color online). Top: Typical contour plot of Eqs.\ (\ref{2v})
for fixed $\bm{J}^{ij}$ and $\bm{n}_j$ ($\bm{J}^{ij}=J(1,.75,-.2)$,
$\theta_j=\pi/3$, $\phi_j=\pi/5$). The two intersection points correspond to
the two solutions $\bm{n}_i^{\pm}$ for $\bm{n}_i$ given by Eq.\ (\ref{ni}).
Bottom: Schematic representation of the $2^{N-1}$ configurations of
$|\Theta\rangle$ compatible with the exchange vectors $\bm{J}^{ij}$ for a spin
chain of $N$ spins and an initial alignment direction at site $j$.}
 \label{f2}
\end{figure}

{\bf Lemma 3.} {\it Given an arbitrary quadratic coupling $\bm{S}_i{\cal
J}^{ij}\bm{S}_j=\sum_{\mu,\nu}J^{ij}_{\mu\nu}S_i^\mu S_j^\nu$ between two spins
and an arbitrary  alignment direction $\bm{n}_j$ of one of the spins, there is
at least one alignment direction $\bm{n}_i$ of the other spin satisfying the factorization Eqs.\
(\ref{1v0}), given by }
\begin{equation}
\bm{n}_i=\alpha[\bm{a}\times\bm{b}\pm(\eta\lambda_+\bm{a}+\lambda_-\bm{b})]\,,
\label{ni} \end{equation}
where $\bm{a}={\cal J}^{ij}\bm{n}^{x'}_j$, $\bm{b}={\cal J}^{ij}\bm{n}^{y'}_j$,   
\begin{equation}
\lambda_{\pm}^2=\frac{\sqrt{(|\bm{a}|^2-|\bm{b}|^2)^2+
		4|\bm{a}\cdot\bm{b}|^2}\pm(|\bm{a}|^2-|\bm{b}|^2)}{2}\,,
\label{lam2}
\end{equation}
and $\alpha$ is a normalization factor, with $\eta=1$ if $\bm{a}\cdot\bm{b}\geq
0$ and $-1$ otherwise (if $\bm{a}\cdot\bm{b}=0$, $\lambda_+$ or
$\lambda_-$ vanishes and the sign of $\eta$ becomes irrelevant). Each sign in
(\ref{ni}) originates a distinct solution for $\bm{n}_i$ (Fig.\ \ref{f2})
if $\lambda_{\pm}$ are not both zero. If $\bm{b}\propto \bm{a}$ and 
$\bm{a}\neq \bm{0}$, then $\bm{n}_i\propto \bm{a}$. Eq.\ (\ref{ni}) holds if $\bm{a}$ 
and $\bm{b}$ are not both $\bm{0}$ (see Appendix for the proof and additional details,  
including the special case $\bm{a}=\bm{b}=\bm{0}$). 

This Lemma implies that at least in open one-dimensional systems of $N$ spins
with {\it arbitrary, not necessarily uniform}  first neighbor quadratic
couplings,  a fully separable eigenstate, compatible with a given (arbitrary)
spin alignment direction of {\it one} of the spins, {\it always exists for suitable
fields at each site}.  The alignment directions of the remaining spins are
determined by successive applications of this Lemma, while  the fields are
determined by Eqs.\ (\ref{ht})--(\ref{hp}). Furthermore, there are typically
 $2^{N-1}$ configurations of spin directions compatible with the couplings
and the initial $\bm{n}_j$, as illustrated in Fig.\ \ref{f2}.

{\bf Lemma 4.} {\it For a pair of equal spins ($S_i=S_j=S$) interacting through
$XYZ$ couplings, if Eqs.\ (\ref{2v}) are satisfied for  non-antiparallel
directions $\bm{n}_i$ and $\bm{n}_j$, there always exist parallel fields
$\bm{h}^{ij}_{\parallel}$ and $\bm{h}^{ji}_{\parallel}$ at $i$ and $j$ such
that the factorizing field $\bm{h}^{ij}_{s}$ for the single pair is uniform:}
\begin{equation}\bm{h}^{ij}_{\parallel}+\bm{h}^{ij}_{\perp}=
\bm{h}^{ji}_{\parallel}+\bm{h}^{ji}_{\perp}=\bm{h}^{ij}_{s}\,.\label{u}\end{equation}

In the uniform case $\bm{n}_i=\bm{n}_j$, it is apparent from Eq.\ (\ref{hp})
that the perpendicular fields are equal, entailing that
$h^{ij}_\parallel=h^{ji}_\parallel$ with the strength $h^{ij}_{\parallel}$
remaining {\it arbitrary}. However, when $\bm{n}_i\neq\bm{n}_j$,  Eq.\
(\ref{u}) leads to fixed values of the parallel fields $h^{ij}_{\parallel}$,
$h^{ji}_{\parallel}$ (explicitly determined in the Appendix) and the pair {\it
uniform} factorizing fields $\bm{h}^{ij}_{s}$ belong to the ellipsoid
($\mu\neq\nu\neq \sigma$)
\begin{equation}
 \sum_{\mu=x,y,z}\frac{({h}^{ij}_{s\mu})^2}{(J_\mu^{ij}+J_{\nu}^{ij})(J_\mu^{ij}
 +J_{\sigma}^{ij})}=S^2\,. \label{Kur}
 \end{equation}
This equation is just that  determined by Kurmann {\it et. al.} in \cite{KU.82}
for the N\'eel-type separable GS in an  antiferromagnetic cyclic chain with
first neighbor couplings in a uniform field. Hence, for a uniform field we
recover this result. Note, however, that under a uniform field  such state will
be  two-fold degenerate if $\bm{n}_i\neq \bm{n}_j$, due to breaking of
permutational symmetry \cite{MRC.15}, and addition of {\it local nonuniform}
parallel fields is necessary to split this degeneracy.

{\bf Lemma 5:} {\it Pairwise entanglement reaches full range in the vicinity
of factorization}.

This result, proved in the Appendix, extends a previous result  shown for
uniform couplings and fields \cite{MRC.15} to the present general case of
non-uniform fields and couplings. It means that pairwise entanglement, though
obviously vanishing at a separable eigenstate, reaches full range if either the
fields or the couplings are slightly varied around the factorization values. It
holds for any number $N$ of spins and any spin $S>0$. Factorization can then be
also considered as an entanglement critical point in the present general
setting.

\section{Examples \label{3}}
As illustration  of the previous lemmas, we discuss  here some special examples
of separable eigenstates and show explicit results for the pairwise
entanglement in the vicinity of the present general factorization conditions.
\subsection{Spin spiral and other separable eigenstates}
We consider  from the present perspective (i.e., starting from the state and
deriving the compatible couplings and fields) three examples of separable
eigenstates : i) constant $\theta$ ($\theta_i=\theta$ in all alignment
directions $\bm{n}_i$),  which includes in particular   {\it spin spiral-type}
eigenstates, ii) constant $\phi$ ($\phi_i=\phi$ $\forall$ $\bm{n}_i$) and iii)
uniform (constant $\theta$ and $\phi$).

i)  Let us first  consider $\theta_i=\theta$ for all
spins, with $\phi_i$ arbitrary. If $\bm{U}^{ij}$, $\bm{V}^{ij}$ are linearly
independent, which implies here that $\bm{n}_i+\bm{n}_j$ does not belong to a
principal plane, Eqs.\ (\ref{2v}) or (\ref{jc2}) lead to an {\it $XXZ$ coupling},
\begin{equation}J_x^{ij}=J_y^{ij}=J^{ij},\;\;J_z^{ij}=J^{ij}\cos(\phi_i-\phi_j)\,,
 \label{phm}\end{equation}
with $J^{ij}$ arbitrary, which is independent of {\it both} $\theta$ and
the average $(\phi_{i}+\phi_j)/2$.
From Eq.\ (\ref{hp}) it can be seen that
 the perpendicular factorizing fields
belong to the principal plane $xy$:
$\bm{h}^{ij}_{\perp}=J^{ij}S_j\sin(\phi_i-\phi_j)(\bm{e}_z\times \bm{n}_i)\,.$

\begin{figure}[t]
 \centerline{\scalebox{.5}{\includegraphics{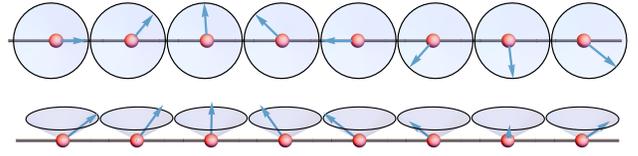}}}
\caption{(Color online). Schematic representation of the spin spiral separable
eigenstate in a spin chain. The alignment direction of the spin at site $i$ is
determined by $\theta_i=\theta$ and $\phi_{i}=\phi_1+(i-1)\Delta\phi$ $\forall$
$i$. The top (bottom) panel corresponds to a top (side) view.}
 \label{f3}
\end{figure}

In particular, considering now a one-dimensional chain with first neighbor
couplings, this case admits solutions with {\it uniform} anisotropy
$J^{ij}_z/J^{ij}=\cos\Delta\phi$, and hence also {\it uniform} couplings  if
$\Delta\phi=\phi_{i+1}-\phi_i$ is {\it constant}. In a cyclic chain we should
have in addition $\Delta\phi=2\pi k/N$, with  $k$ any integer between $1$ and
$N-1$,  as schematically shown in Fig.\ \ref{f3}. For a spin $S$ chain the
total perpendicular factorizing fields
$\bm{h}^{i}_{\perp}=\bm{h}^{i,i-1}_{\perp}+\bm{h}^{i,i+1}_{\perp}$ become
\begin{eqnarray}
\bm{h}^i_{\perp}&=&(J^{i-1,i}-J^{i,i+1})S\sin\Delta\phi(\bm{e}_z\times\bm{n}_i)\,.
\label{hsw}
\end{eqnarray}
Eq.\ (\ref{hsw}) shows that for {\it uniform} couplings ($J^{i,i+1}=J$ $\forall
i$), $\bm{h}^{i}_{\perp}=\bm{0}$ and {\it the spin spiral is an exact
eigenstate of the present $XXZ$ cyclic chain already at zero field} (in the
open case, the endpoint fields $\bm{h}^{1}_{\perp}$ and $\bm{h}^{N}_{\perp}$
remain non-zero). In the cyclic case it  corresponds to  a highly degenerate
eigenvalue of $H$ that arises when $J_z^{ij}/J^{ij}=\cos \frac{2\pi k}{N}$.  In
the presence of parallel fields $\bm{h}_{\parallel}^i=h_{\parallel}\bm{n}_i$,
the degeneracy will be removed, its energy becoming
\begin{equation}
E_{\Theta}=-NS(h_{\parallel}+JS\cos \Delta\phi) \,.
\label{esw}
\end{equation}
It will then be a nondegenerate GS if $h_{\parallel}$ is sufficiently large
(typically $h_{\parallel}=O(|J|S))$.

ii) Let us consider now $\phi_i=\phi$ for all spins,
with  the angles $\theta_i$ remaining arbitrary. Assuming again $\bm{U}^{ij}$ and
$\bm{V}^{ij}$ linearly independent, i.e.\ that  $\bm{n}_i+\bm{n}_j$
does not belong to a principal plane, Eq.\ (\ref{jc2}) leads again to an {\it
$XXZ$-type coupling},
\begin{equation}
J_x^{ij}=J_y^{ij}=J^{ij}(1-\eta_{ij}^2),\;\;J_z^{ij}=J^{ij}(1+\eta_{ij}^2)\,,
 \label{thm}\end{equation}
where $\eta_{ij}=\sin(\theta_j-\theta_i)/\sin\bar{\theta}_{ij}$, with
$\bar{\theta}_{ij}=(\theta_i+\theta_j)/2$ and $J^{ij}$ arbitrary.  Hence, the
coupling  is {\it independent of $\phi$} but depends now on $\theta_i$,
$\theta_j$, with 
$|J^{ij}_z|\geq |J^{ij}|$. For small $\theta_i-\theta_j=\delta\theta$, the orthogonal fields are
$\bm{h}^{ij}_{\perp}\approx
-J^{ij}S_j\delta\theta\cos\bar{\theta}_{ij}(\cos\phi,\sin\phi,-\tan\bar{\theta}_{ij})$,
which  belongs to the plane defined by $\bm{n}_i$ and $\bm{n}_j$. In a
one-dimensional chain with first neighbor couplings, a constant coupling
becomes feasible for a N\'eel-type configuration with alternating angles
$\theta_1\theta_2\theta_1\ldots$, since in this case $\bar{\theta}_{ij}$ and
$|\theta_i-\theta_j|$ are constant. The energy $E_\Theta$ is in any case
independent of $\phi$, with $\langle \bm{S}_i\rangle\cdot{\cal
J}^{ij}\langle\bm{S}_j\rangle
=S_iS_jJ^{ij}[\eta_{ij}^2\cos2\bar{\theta}_{ij}+\cos(\theta_i-\theta_j)]$.

iii) Let us finally consider a fixed alignment direction
$\bm{n}_i=\bm{n}$ for all spins ($\theta_i=\theta$, $\phi_i=\phi$ $\forall$
$i$). If $\bm{n}$ does not belong to a principal plane, $\bm{U}^{ij}$ and
$\bm{V}^{ij}$ are linearly independent and Eqs.\ (\ref{phm}) or (\ref{thm})
lead to $J_\mu^{ij}=J^{ij}$, i.e., to an {\it isotropic} coupling
$\propto\bm{S}_i\cdot\bm{S}_j$. Eq.\ (\ref{hp2}) then implies
$\bm{h}^{ij}_{\perp}=\bm{0}$, i.e.\ no orthogonal field is required since such
uniform state is already an obvious eigenstate of $\bm{S}_i\cdot\bm{S}_j$ for
{\it any} orientation $\bm{n}$.

If $\bm{n}$  belongs instead to a principal plane $\mu\nu$ ($n_\sigma=0$),
$\bm{V}^{ij}=\bm{0}$ and anisotropic couplings become also feasible,
provided $\bm{J}$ is orthogonal to $\bm{U}^{ij}$. This condition leads to
\begin{equation}
J^{ij}_\sigma=J_\mu^{ij}n_{\nu}^2+J_\nu^{ij}n_{\mu}^2
=J^{ij}_\nu+(J^{ij}_\mu-J^{ij}_\nu)\cos^2\gamma\,, \label{jp2}
\end{equation}
with $J^{ij}_\mu$, $J^{ij}_\nu$ {\it arbitrary} and $\gamma$ the angle between
$\bm{n}$ and the $\mu$ axis, implying a fixed ratio
$\frac{J^{ij}_\sigma-J^{ij}_\nu}{J^{ij}_\mu-J^{ij}_\nu}$ \cite{MRC.15}. The
factorizing fields  belong to the same principal plane, with
\begin{equation}
\bm{h}^{i}_{\perp}=\sin\gamma \cos\gamma
(\bm{e}_\sigma\times\bm{n})\sum_j S_j(J^{ij}_\mu-J^{ij}_\nu)\,.\label{hu}
\end{equation}

\subsection{Pairwise Entanglement}

\begin{figure}[h!]
 \centerline{\scalebox{.8}{\includegraphics{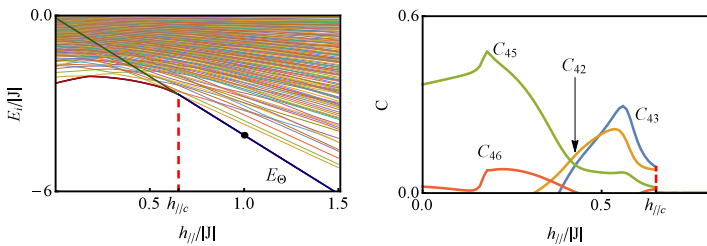}}}
\caption{(Color online) Left panel: Scaled energy spectrum of a finite spin-$1/2$
chain with first neighbor $XYZ$ couplings and 8 spins when a factorizing field
$\bm{h}^i=\bm{h}^i_\perp + h_\parallel\bm{n}^i$ with $\bm{h}^i_\perp$ fixed and
$h_\parallel\geq 0$ is applied. The decreasing straight line represents the
energy $E_\Theta$ of the separable eigenstate $|\Theta\rangle$, which
becomes GS for $h_\parallel>h_{\parallel c}$ (dashed line). Right panel: GS
concurrences $C_{ij}$  between a central spin and first and second neighbors,
showing that for  $h_\parallel<h_{\parallel c}$ the GS is entangled whereas for
$h_\parallel>h_{\parallel c}$ it is completely separable. All labels dimensionless.}
 \label{f4}
\end{figure}

\begin{figure}[h!]
 \centerline{\scalebox{.75}{\includegraphics{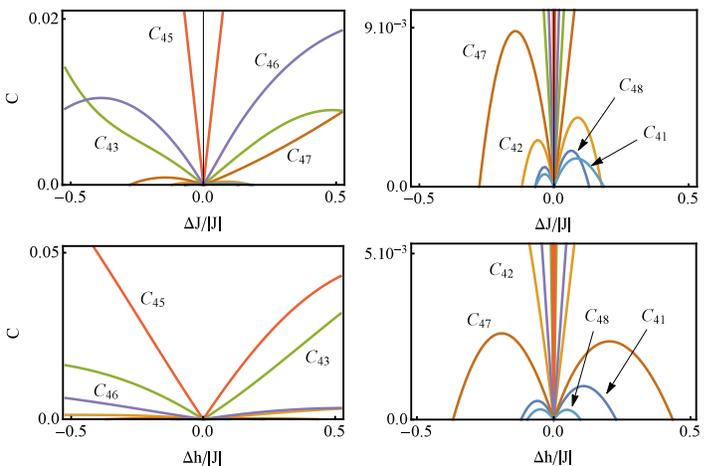}}}
\caption{(Color online) GS concurrences $C_{ij}$ between spins $i$ and $j$ in
the chain of Fig.\ \ref{f4} in the vicinity of factorization (point at
$h_{\parallel}/|\bm{J}|=1$ on Fig.\ \ref{f6}). In the top panel, we have set
the couplings as $J^{ij}_\mu+\Delta J$ $\forall$ $i,j$, with $J^{ij}_\mu$ the
factorizing values, such that $C_{ij}=0$ $\forall$ $i,j$ if $\Delta J=0$. In
the bottom panel, the couplings are fixed at the factorizing values but the
fields are now $\bm{h}^i+\Delta h \bm{n}_{i\perp}$, with
$\bm{n}_{i\perp}\propto\bm{h}^i_{\perp}$,  such that $C_{ij}=0$ $\forall$ $i,j$
at $\Delta h=0$. $C_{ij}$ is verified to reach full range in the vicinity of
the present general factorization point (nonuniform couplings and fields).
Right panels show the same quantities of the left panels at a smaller scale.
All labels dimensionless.}
 \label{f5}
\end{figure}

We now show in Figs.\ \ref{f4}--\ref{f5} the behavior of pairwise entanglement
in the GS of a finite spin-$1/2$ chain with non-uniform first neighbor couplings under
non-uniform fields.  The  entanglement between
spins $i$ and $j$ is measured through the concurrence \cite{W.98}
$C_{ij}=2\lambda_{\rm max}-{\rm Tr}\,M_{ij}$ where $\lambda_{\rm max}$ is the
largest eigenvalue of
$M_{ij}=[\rho_{ij}^{1/2}\tilde{\rho}_{ij}\rho_{ij}^{1/2}]^{1/2}$, with
$\tilde{\rho}_{ij}=\sigma_y\otimes\sigma_y\rho^*_{ij}\sigma_y\otimes\sigma_y$
in the standard basis and $\rho_{ij}$ the reduced state of spins $i$ and $j$.

We consider a completely separable eigenstate state with the spin alignment
directions of the spins  selected at {\it random}. The exchange couplings
between every adjacent pair were then obtained through Eq.\ (\ref{jc}), setting
a uniform norm  $|\bm{J}^{ij}|=J$  for all exchange vectors. In order for
$|\Theta\rangle$ to be a GS, nonuniform fields $\bm{h}^i= \bm{h}^i_\perp +
h_\parallel\bm{n}^i$ with $\bm{h}^i_\perp$ the fixed orthogonal factorizing
fields (\ref{hp}) and $h_\parallel\geq 0$, were applied at each site. At
$h_\parallel=0$ $|\Theta\rangle$ is an exact eigenstate of $H$ although not
the GS. As shown on the left panel of Fig.\ \ref{f4}, the energy
$E_\Theta$, given by Eq.\ (\ref{EE2}), decreases linearly (and with maximum
slope) for increasing $h_\parallel$, and at  $h_\parallel=h_{\parallel c}$ a GS
transition occurs, such that $|\Theta\rangle$ becomes GS $\forall$
$h_\parallel>h_{\parallel c}$. Accordingly, GS pairwise concurrences $C_{ij}$
vanish for $h_\parallel>h_{\parallel c}$ $\forall$ $i,j$, as seen on the right
panel of Fig.\ \ref{f4}.

The behavior of pairwise concurrence in the vicinity of
factorization is shown on Fig.\ \ref{f5}. We have chosen a stable factorized
GS (the point at $h_{\parallel}/|\bm{J}|=1$ in Fig.\ \ref{f4}), such that the
perturbations considered led to a smooth variation of the GS, without crossings
with the first excited state. The correction to the field was chosen
perpendicular to the alignment direction as any perturbation $\Delta
h_{\parallel}  \bm{n}_{i}$ just shifts the GS energy. It is verified that all
concurrences $C_{ij}$ are turned on in the immediate vicinity of factorization
for variations of the couplings (top panel) or fields (bottom panel),
although those for distant pairs can be very small and vanish outside a small
interval. Nonetheless, the factorization point stands out as an entanglement
``critical point'' of the system, in the sense of exhibiting infinite range in 
its vicinity. Note also that coefficients $\beta_{ij}$ in the  reduced state
of the pair, Eq.\ (\ref{red}), will vanish and hence
change their signs at $\Delta J=0$ or $\Delta h=0$.

\section{Separable ground state engineering\label{4}}
One of the goals of this paper is to provide recipes for engineering
nondegenerate maximally aligned exactly separable GS in spin systems. In the
previous section this problem was approached from two different perspectives:
I) Specifying the alignment directions $\bm{n}_{i}$ of the spins and finding
compatible exchange vectors $\bm{J}^{ij}$ (Lemma $2$), and  II) Assuming fixed
exchange couplings $\bm{J}^{ij}$ and finding compatible alignment directions of
the spins (Lemma $3$). The first scheme has, for instance, enabled to easily
identify spin-spiral type separable eigenstates in $XXZ$ chains with
special values of $J_z/J_x$, already at zero field.

In the first case it is evident that a necessary condition for engineering the
separable GS is that the exchange coupling between the spins must be {\it
tunable}. This could in principle be feasible in spin systems based on quantum
dots \cite{VS.15,BB.15}, superconducting Josephson junctions \cite{TK.02} and
nuclear (or electron-nuclear) spin states \cite{CW.04}.  In the second scenario
the exchange couplings are {\it fixed} and Lemma $3$ yields the possible
separable eigenstates the system can posses. This is a more restrictive case,
as we suppose little (to none) control over the exchange couplings. Thus,
according to how much control is available over the system the problem may be
considered from one standpoint or the other. To quantify such control (and
assuming  that a uniform field can always be applied) we introduce the
{\it experimental complexity}  ``$\bm{\varepsilon_c} =\bm{(m,k)}$'', which
indicates that for a system of $N$ interacting spins to have a given separable
state as its non-degenerate GS, control over $m\leq N-1$ local fields and $k$ exchange
couplings between spins is required. As expected, the separable state which
requires the simplest control will be shown to be the uniform separable state.

\subsection{Tunable exchange couplings}

\begin{figure}[h!]
 \centerline{\scalebox{1.}{\includegraphics{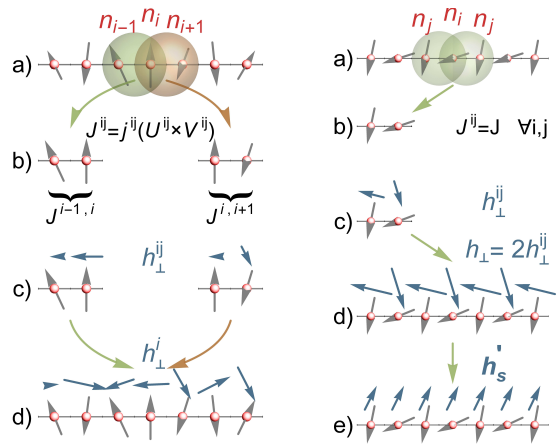}}}
\caption{(Color online). Schematic representation of the separable state with
arbitrary spin alignment directions  $\bm{n}_i$ at each site (left panel) and
with a N\'eel-type configuration  (right panel). From top to bottom: a) The
alignment directions at each site are specified. Then, for each pair the b)
exchange couplings and c) perpendicular fields are determined. d) The
perpendicular factorizing field at each site is
$\bm{h}^i_{\perp}=\sum_j\bm{h}^{ij}_{\perp}$.
e) The uniform factorizing field
$\bm{h}'_s$ is determined  by Lemma $4$.}
 \label{f6}
\end{figure}

As shown in Lemma $2$,  by specifying the individual spin alignments $\bm{n}_i$
and $\bm{n}_j$ of an interacting pair, the exchange vector $\bm{J}^{ij}$ and
the fields $\bm{h}_\perp^{ij}$, $\bm{h}_\perp^{ji}$ can be determined. Then, by
applying suitable parallel fields at each site, the separable state
$|\Theta\rangle$ can be made a nondegenerate GS of $H$ (Lemma 1). Assuming that
a uniform field can be applied, then $\bm{\varepsilon_c} =\bm{(1,1)}$ for a
single pair. Similarly, for a chain of $N$ spins with first neighbor couplings,
$\bm{\varepsilon_c}=\bm{(N-1,N-1)}$ in the open case and
$\bm{\varepsilon_c}=\bm{(N-1,N)}$ in the cyclic case (Fig. \ref{f6}).

A possible way of reducing the complexity is to obtain separability by only
applying a uniform factorizing field. According to Lemma $4$, this is always
possible in a system of two identical spins if Eqs.\ (\ref{2v}) are fulfilled,
provided the alignment directions $\bm{n}_1$ and $\bm{n}_2$ are not
antiparallel. Nonetheless, if $\bm{n}_1\neq \bm{n}_2$ such eigenstate will be
two-fold degenerate (due to basic permutational symmetry-breaking) and local
nonuniform parallel fields must be added to split this degeneracy.  Therefore,
the complexity remains unchanged if a nondegenerate GS is to be achieved.  The
same holds for finite cyclic chains with first neighbor couplings and an even
number of spins  if an alternating N\'eel-type separable eigenstate
($\bm{n}_i=\bm{n}_1$ ($\bm{n}_2$) for $i$ odd (even)) is sought. With the same
previous scheme  (and just doubling the field at each site due to the
contributions from each neighbor) it is possible to obtain such eigenstate by
applying a uniform field (right panel in Fig.\ \ref{f6}). This state can be a
GS for antiferromagnetic-type couplings \cite{KU.82}, although it will be again
degenerate. An additional alternating field will then be required to turn it
into a nondegenerate GS.

On the other hand, if $S_i=S_j$ and $\bm{n}_i=\bm{n}_j$, the uniform
factorizing field is just $\bm{h}_{s}=\bm{h}_{\parallel}+\bm{h}_{\perp}$ with
the strength $h_{\parallel}$ remaining arbitrary. Thus, according to Lemma $1$
it is possible to make this state a nondegenerate GS with an arbitrarily large
spectral gap by applying just a uniform field $\bm{h}_{s}$, implying
$\bm{\varepsilon_c} =\bm{(0,1)}$. Similarly, in a cyclic chain of $N$ (even or
odd) spins with first neighbor couplings, such states require
$\bm{\varepsilon_c}=\bm{(0,N)}$, due to the same arguments. In an open chain it
is necessary, however, to correct the fields at the borders due to one missing
neighbor and hence $\bm{\varepsilon_c}=\bm{(2,N-1)}$.

Achieving the necessary control over the exchange interactions and local
magnetic fields is a challenge in itself. However, this requirement can be
relaxed by considering spin clusters  schemes \cite{MJD.03} where the qubit is
encoded in several spins. The previous schemes can then be used as building
blocks to engineer (bulk per bulk) the separable GS. If the spin configuration
of each cluster is uniform,  the factorizing fields at each bulk will also be
uniform, and we would only require control over the exchange couplings and
fields at the border of the clusters.

\subsection{Fixed exchange couplings}
\begin{figure}[h!]
 \centerline{\scalebox{1.2}{\includegraphics{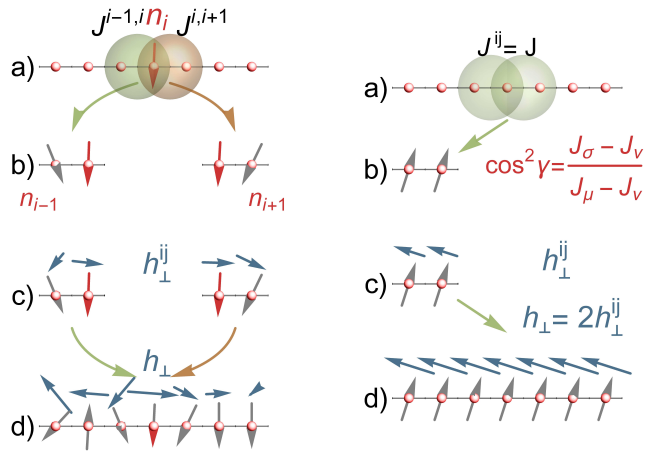}}}
\caption{(Color online). Schematic representation of  a spin chain with
nonuniform couplings (left panel) and with uniform anisotropic couplings
leading to a uniform separable state (right panel). From top to bottom: a) The
exchange couplings are fixed. In the left panel one of the spin alignment
directions $\bm{n}_i$ is specified. b) Left: Using Lemma $3$ the spin alignment
directions for the spins at sites $i\pm1$ can be determined. Right: $\bm{n}$
belongs to the principal plane $\mu\nu$ determined by the exchange couplings.
c-d) Left: By successive applications of Lemma $3$ the spin alignments of  the
remaining spins are determined.}
 \label{f7}
\end{figure}

For a pair of interacting spins, given the alignment direction $\bm{n}_j$ of
one of the spins, according to Lemma $3$ an alignment direction
$\bm{n}_i$ of the remaining spin can always be determined, regardless of the
coupling between them. Then, by appropriate fields the ensuing separable state can be
made an exact GS with $\bm{\varepsilon_c}=\bm{(1,0)}$. In finite arrays Lemma
$3$ can therefore be used to determine spin configurations compatible with the
fixed exchange couplings. For instance, in an open chain of $N$ spins with
first neighbor couplings, by specifying the alignment direction of only one
spin, this method determines the possible alignment directions of the remaining spins
(typically $2^{N-1}$ configurations, Fig.\ \ref{f2}, right panel).
This scheme is represented in the left panel of Fig.\ \ref{f7}. In this case
$\bm{\varepsilon_c}=\bm{(N-1,0)}$, whereas in the cyclic case control on
one exchange coupling is required, meaning that
$\bm{\varepsilon_c}=\bm{(N-1,1)}$.

If in the previous system the exchange couplings are {\it uniform},
$\bm{J}^{ij}=\bm{J}$, the  uniform separable solution  $\bm{n}_i=\bm{n}$
$\forall$ $i$ is {\it always} feasible provided $\bm{n}$ is appropriately
chosen. If the coupling is {\it isotropic}, $J_\mu=J$, $\mu=x,y,z$, then, as
discussed in the previous section, $\bm{n}$ is arbitrary, i.e., the solution
for $\bm{n}_i$ given by Lemma $3$ is the same $\bm{n}_j$ of the initial spin
for any $\bm{n}_j$ (see Appendix). However,  if the exchange interaction is
{\it anisotropic}, a uniform solution is feasible provided $\bm{n}_j$ belongs
to a principal plane and satisfies Eq.\ (\ref{jp2}), as depicted in the right
panel of Fig.\ \ref{f7}. In cyclic chains (with either isotropic or anisotropic
couplings) such uniform $|\Theta\rangle$ can then be made a nondegenerate GS
with just a uniform magnetic field, i.e., $\bm{\varepsilon_c} =\bm{(0,0)}$,
while in open chains $\bm{\varepsilon_c} =\bm{(2,0)}$ due to the border
corrections. On the other hand, N\'eel-type solutions, also feasible for
uniform first neighbor couplings, require an additional alternating field in
order to become a nondegenerate GS. The uniform solution is also directly
feasible for higher range couplings \cite{MRC.15}, as well in more general
arrays and geometries. Just the fields near the border should be adequately
corrected.  The uniform separable GS is therefore that requiring the least
control over the system.

\subsection{Experimental feasibility and implementations}
The possibility of simulating interacting spin systems enables the
aforementioned engineering methods to be realized. We mention
two physical realizations, in which, with the current state of
technology, couplings and fields can be controlled.

{\it  Superconducting devices.} Superconducting qubits based on Josephson
junctions in solid state electrical circuits present an attractive scenario for
a quantum simulator due to their robustness, long coherence times and intrinsic
low dissipation \cite{DS.13}. It has been shown that superconducting flux
qubits circuits can be used (under specific regimes) to simulate $XX$ spin
systems with nearest-neighbor interactions with nontransverse nonuniform fields
(belonging to the $xz$ principal plane of the couplings) \cite{ZI.14}. In these
systems, the exchange couplings  can be tuned if direct current superconducting
quantum interference device (SQUID) are used to couple the flux qubits
\cite{vP.07}, whilst the direction and strength of the local magnetic fields
are controlled by the phases and amplitudes, respectively, of microwave driving
fields. Realizations of chains with first neighbor tunable $XZ$ couplings, not
necessarily uniform, in  nontransverse and nonuniform fields have also been
recently reported using planar transmon qubits (a type of superconducting
charge qubits) \cite{B.16}.

{\it  Trapped Ions.} When interacting with lasers, trap ions can simulate $XYZ$
effective spin systems in nontransverse magnetic fields. Trapped ions methods
are highly controllable, versatile and present long decoherence times and high
readout precision. When simulating interacting spin systems, the exchange
couplings and the effective magnetic fields can be tuned by controlling the
lasers acting on the internal transition of the ions \cite{PC.04,SR.15}.

In these systems, once the nondegenerate separable GS is obtained it can be used in actual 
computing instances which require an initial fully separable state, as in standard models 
for quantum computation and quantum simulation \cite{NC.00,VC.05,GA.14,SR.15}. 
In particular, in typical quantum annealing, one starts from a known separable GS of a trivial 
noninteracting Hamiltonian (usually $h\sum_i S_i^x$)  which is then continuously driven to a target 
interacting Hamiltonian whose nontrivial GS is sought \cite{DC.08,MJ.11} (normally an Ising type 
Hamiltonian $h'\sum_i S_i^z+\sum_{i,j}J^{ij} S_i^zS_j^z$). Thus, the previous factorization schemes 
enable to think of annealing protocols with always-on interactions in which just a part of the original 
Hamiltonian is quenched. For instance, one could start from a convenient maximally aligned uniform 
separable GS in an $XYZ$ system with a nontransverse field, such that the GS is well gapped, and 
continuously decrease the field along the $x$-axis and the $XY$ terms leaving the sought Ising Hamiltonian. 
Additionally, quantum annealing could be used in principle to obtain the GS of a complex Hamiltonian 
starting from a separable GS by an analogous process (i.e., just modifying the fields, initially at 
suitable factorizing values). In particular, in chains with {\it fixed} arbitrary first neighbor couplings, 
Lemma 3 and Lemma 1 ensure the existence of (multiple) completely separable gapped GS if the fields are 
adequately tuned, entailing that the previous annealing could always be applied.

\section{Conclusions\label{5}}
We have presented a novel approach to the problem of factorization, showing the
possibility of starting from the state and determining the compatible couplings
and fields. This approach  opens the way to separable ground state engineering
in interacting spin systems if some control over the couplings (assumed
quadratic) and fields is feasible. For a fully separable state with {\it
arbitrary} spin alignment directions at each site, nonzero couplings between
any selected pairs (which can be all pairs or just some pairs) and finite
fields at each site always exist such that the ensuing Hamiltonian has such
state as a nondegenerate GS. In this way, some noticeable separable eigenstates
(like the spin-spiral solution) were easily identified in $XXZ$ chains. In
addition, in open one-dimensional systems of $N$ spins with arbitrary first
neighbor couplings, at least one (and typically many) exactly separable GS
compatible with an arbitrary spin direction at one site is always  feasible if
the fields can be tuned at each site. And for a single pair, the field can
always be chosen as uniform.  Furthermore, pairwise entanglement reaches full
range in the immediate vicinity of factorization (for perturbations in the
fields or couplings), regardless of the type of solution, indicating that even
in the present general setting factorization can still be considered as an
entanglement critical point. These results, while providing useful insight into
interacting spin systems and models without analytical solution, enable to
devise separable ground state engineering methods which could be used in
quantum information protocols and quantum annealing. 

\begin{acknowledgments}
The authors acknowledge support from CONICET (MC, NC) and CIC (RR) of Argentina.
Work supported by CONICET PIP 112201101-00902.
\end{acknowledgments}

\appendix*
\section{}
We present here proof details of the Lemmas presented in section \ref{lem}.

{\bf Lemma 1.} As previously stated, there is no state $|\Psi\rangle$
orthogonal to $|\Theta\rangle$ whose energy decreases more rapidly with
$h^i_{\parallel}$ than $E_{\Theta}$. For instance, if $S_i=S$ $\forall$ $i$,
$|\langle\bm{S}_i\cdot {\cal J}^{ij}\bm{S}_j\rangle_{\Psi}|\leq J^{ij}S(S+1)$
is bounded (with $J^{ij}$ the maximum singular value of the matrix ${\cal
J}^{ij}$) while
$\langle-\sum_i\bm{S}_i\cdot\bm{h}^i\rangle_{\Psi}\geq-(N-1)Sh_{\parallel}
-\sum_i\langle\bm{S}_i\rangle_{\Psi}\cdot\bm{h}^i_{\perp}$ if
$\bm{h}^i=h_{\parallel}\bm{n}_i+\bm{h}^i_{\perp}$ and $N$ is the number of
spins. Hence, no state $|\Psi\rangle$ can match the decrease with increasing
$h_{\parallel}$ of $E_{\Theta}$,  which will contain a term
$\propto-NSh_{\parallel}$. Typically, if all $J^{ij}$ are $O(J)$, the threshold
parallel field $h_{\parallel\,c}$ for making $|\Theta\rangle$ a GS will be
$O(JSl)$, with $l$ the number of  neighbors coupled with a given spin.\\

{\bf  Lemma 2.} If  ${\bm J}^{ij}$ is {\it orthogonal} to the subspace spanned
by the vectors $\bm{U}^{ij}$ and $\bm{V}^{ij}$,  Eqs.\  (\ref{2v}) are satisfied. Here we
explicitly determine these exchange couplings and the concomitant factorizing
fields for the two different situations that need to be considered:

{\bf a)} {\it $\bm{U}^{ij}$ and $\bm{V}^{ij}$ linearly independent}. In this case
 $\bm{J}^{ij}$ is given by Eq.\ (\ref{jc}), which can be explicitly written
in terms of the alignment directions $\bm{n}_i,\bm{n}_j$ as
\begin{equation}
J^{ij}_\mu=-j^{ij} (n_{j\mu}D_i/n_{i\mu}+n_{i\mu}D_j/n_{j\mu})\,,
\label{jc2}
\end{equation}
where $ D_i=\prod_\mu n_{i\mu}=\frac{1}{4}\sin\theta_i\sin 2\theta_i\sin
2\phi_i$. It satisfies $J^{ij}_\mu=J^{ji}_\mu$.
The orthogonal fields $\bm{h}^{ij}_{\perp}$ can then be specified just
in terms of the alignment directions and the constants $j^{ij}$:
\begin{equation}
(\bm{h}^{ij}_{\perp})_\mu=j^{ij}S_j(n_{j\mu}^2-n_{i\mu}^2)D_i/n_{i\mu}\,.\label{hp2}
\end{equation}
Correspondingly, the energy $E_{\Theta}$, (\ref{EE2}),  becomes
\begin{equation}
E_{\Theta}=-\sum_i S_i h^i_{\parallel}-{\textstyle\frac{1}{2}}
\sum_{i\neq j}j^{ij}S_i S_j(D_i+D_j)\,.
\label{E1}
\end{equation}

{\bf b)} {\it $\bm{U}^{ij}$ and $\bm{V}^{ij}$ linearly dependent}.
For the present orthogonal vectors $\bm{n}_i^{\mu'}$, this case occurs when 
${\bm V}^{ij}=\bm{0}$ (and $\bm{U}^{ij}\neq\bm{0})$. Hence,
Eqs.\ (\ref{2v}) define a {\it plane} orthogonal to $\bm{U}^{ij}$
of exchange vectors $\bm{J}^{ij}$:

i) If  $\bm{n}_i$ and $\bm{n}_j$ belong to the same principal plane, say
$\mu\nu$, with $\sigma$ the direction orthogonal to this plane
($n_{i\sigma}=n_{j\sigma}=0$), Eqs.\ (\ref{2v}) lead to
\begin{equation}
J^{ij}_\sigma=J_\mu^{ij}n_{i\nu}n_{j\nu}+J_\nu^{ij}n_{i\mu}n_{j\mu}\,, \label{jp}
\end{equation}
with $J_\mu^{ij}$ and $J_\nu^{ij}$ {\it arbitrary}. There are now {\it two}
independent exchange couplings, which are those of the plane containing the
alignment directions (bottom right panel in Fig.\ \ref{f1}). From Eq.\
(\ref{hp}) it is seen that $\bm{h}_{\perp}^{ij}$ also belongs to the principal
plane $\mu\nu$, directly depending  on the free couplings $J^{ij}_\mu$ and
$J^{ij}_\nu$. Moreover, by choosing them such that
$\bm{n}_i\times(\bm{J}^{ij}*\bm{n}_j)=\bm{0}$, then
$\bm{h}_{\perp}^{ij}=\bm{0}$, i.e., $|\Theta\rangle$ is an exact eigenstate at
zero field.

ii) If  $\bm{n}_i$ is the reflection of $\bm{n}_j$ with respect to the
principal plane $\mu\nu$, with all components of $\bm{n}_{i(j)}$ nonzero
(otherwise we return to previous case i)  Eqs.\ (\ref{2v}) lead to
\begin{equation}
J^{ij}_\sigma={\frac{J_\mu^{ij}(1-n_{i\mu}^2)+J_\nu^{ij}(1-n_{i\nu}^2)}{1-n_{i\sigma}^2}}\,,
\label{jnus}
\end{equation}
where $J_\mu^{ij}$ and $J_\nu^{ij}$ are arbitrary.
Then, from Eq.\ (\ref{hp}) the orthogonal fields are
\begin{eqnarray}
(\bm{h}^{ij}_{\perp})_\sigma&=&S_j(J_\mu^{ij}+J_\nu^{ij})n_{i\sigma}\,,
\nonumber\\(\bm{h}^{ij}_{\perp})_{\mu(\nu)}&=&
S_j(J_{\nu(\mu)}^{ij}-J_\sigma^{ij})n_{i\mu(\nu)}\,.
\label{hnus}
 \end{eqnarray}
iii) Finally, if $\bm{n}_i=-\bm{n}_j$, we should just replace $J_{\sigma}^{ij}$
by $-J_{\sigma}^{ij}$ in (\ref{jnus})--(\ref{hnus}),
 such that $\bm{J}^{ij}$  belongs to the plane
 $\sum\limits_{\mu=x,y,z}\!\!\!J_\mu^{ij}(1-n_\mu^2)=0$,
with $(\bm{h}^{ij}_{\perp})_\mu=S_j({\rm Tr}({\cal J})-J_\mu)n_{i\mu}$.

A final remark is that if one approaches any of the cases {\bf b)} from the
linearly independent case {\bf a}), it is verified that all previous equations
(\ref{jp})--(\ref{hnus}) are in agreement with the corresponding limit of Eqs.\
(\ref{jc2})--(\ref{hp2}). \qed \\

{\bf Lemma 3.} {\it Proof:} Assuming first $\bm{a}={\cal J}^{ij}\bm{n}^{x'}_j$ and 
 $\bm{b}={\cal J}^{ij}\bm{n}^{y'}_j$ linearly independent, we can define
orthonormal vectors $\bm{k}$, $\bm{l}$ such that $\bm{a}=|\bm{a}|\bm{k}$,
$\bm{b}=b_1\bm{k}+b_2\bm{l}$, with $b_1=\bm{a}\cdot\bm{b}/|\bm{a}|$.  
Then normalized vectors $\bm{n}_i^{x'}\propto
b_2\bm{k}-b_1\bm{l}+\lambda\bm{m}$ and $\bm{n}_i^{y'}\propto
\lambda\bm{l}+b_1\bm{m}$, with $\bm{m}=\bm{k}\times \bm{l}$, satisfy
$\bm{n}_i^{x'}\cdot\bm{b}=\bm{n}_i^{y'}\cdot\bm{a}=\bm{n}_i^{x'}\cdot\bm{n}_i^{y'}=0$. 
Hence, the factorization  Eqs.\ (\ref{1v0}) are fulfilled provided 
$\bm{n}_i^{x'}\cdot\bm{a}=\bm{n}_i^{y'}\cdot\bm{b}$, which implies
$\lambda=\pm\lambda_+$, with $\lambda$ given by Eq.\ (\ref{lam2}). 
A suitable alignment direction at site $i$ can then be obtained as
$\bm{n}_i=\bm{n}_i^{x'}\times \bm{n}_i^{y'}$, which yields Eq.\ (\ref{ni}). 

Additionally, if $\bm{b}\propto \bm{a}$, with $\bm{a}\neq \bm{0}$, Eq.\
(\ref{ni}) still holds, since in this case it leads to 
$\bm{n}_i\propto\bm{a}$,  which  is indeed an obvious solution for $\bm{n}_i$ of Eqs.\ (\ref{1v0}). And if
$\bm{a}=\bm{b}=\bm{0}$, which occurs iff both $\bm{J}^{ij}$ (assumed non-zero)
and $\bm{n}_j$ point along the same principal axis ($\mu$) then $\bm{n}_i$
remains {\it arbitrary}. The effect of the coupling on the product state can
here be balanced by a factorizing field, as it involves just one-spin
excitations: $J^{ij}_\mu S_i^\mu S_j^\mu|\Theta\rangle=S_j J^{ij}_\mu
S_i^\mu|\Theta\rangle$. \qed

As is evident from Eq.\ (\ref{ni}),  two different solutions for ${\bm n}_i$ exist unless
$\lambda_\pm$ are simultaneously zero. This case arises, for instance, if
$\bm{J}^{ij}\propto (1,1,1)$ (isotropic coupling) or if all components of
$\bm{J}$ have the same absolute value (e.g. $\bm{J}\propto (1,1,-1)$), which
imply $|\bm{a}|=|\bm{b}|$ and  $\bm{a}\cdot\bm{b}=0$ in
(\ref{lam2}).  In the isotropic case,
$\bm{a}=\bm{n}_j^{x'}$, $\bm{b}=\bm{n}_j^{y'}$ and Eq.\ (\ref{ni}) implies then 
the single solution $\bm{n}_i=\bm{n}_j$ (uniform solution).\\

{\bf Lemma 4.} 
{\it Proof:} From Eq.\ (\ref{hp}), $\bm{h}^{ij}_{\perp}=-S[{\cal
	J}^{ij}\bm{n}_j-\bm{n}_i(\bm{n}_i\cdot{\cal J}^{ij}\bm{n}_j)]$ and Eq.\
(\ref{u}) implies
\begin{eqnarray}
S\bm{J}^{ij}\ast(\bm{n}_j-\bm{n}_i)&=&[h^{ij}_\parallel+
S(\bm{n}_i\cdot\bm{J}^{ij}\ast\bm{n}_j)]\bm{n_i}-\nonumber\\
&&[h^{ji}_\parallel+S(\bm{n}_j\cdot\bm{J}^{ij}\ast\bm{n}_i)]
\bm{n_j}\,, \label{d1}
\end{eqnarray}
which is verified for some $h^{ij}_\parallel$ and $h^{ji}_\parallel$ iff
$\bm{J}^{ij}\ast(\bm{n}_j-\bm{n}_i)$ belongs to the subspace generated by
$\bm{n}_i$ and $\bm{n}_j$. If $\bm{n}_i=\bm{n}_j$
this condition is trivially satisfied (with
$h^{ij}_{\parallel}=h^{ji}_{\parallel}$, arbitrary)  while if $\bm{n}_i$ and $\bm{n}_j$
are not collinear,  this condition implies
$(\bm{J}^{ij}\ast(\bm{n}_i-\bm{n}_j))\cdot(\bm{n}_i\times\bm{n}_j)=0$, i.e.,
$\bm{J}^{ij}\cdot[(\bm{n}_i-\bm{n}_j)\ast(\bm{n}_i\times\bm{n}_j)]=0.$ But this
equation is always fulfilled if $\bm{J}^{ij}\propto
\bm{U}^{ij}\times\bm{V}^{ij}$ (Eq.\ (\ref{jc})), while if $\bm{U}^{ij}$ and
$\bm{V}^{ij}$ are linearly dependent, it is fulfilled by {\it any}
$\bm{J}^{ij}$, since in this case
$(\bm{n}_i-\bm{n}_j)\ast(\bm{n}_i\times\bm{n}_j)=\bm{0}$. No solution exists,
however, if $\bm{n}_i=-\bm{n}_j$. 

In the last antiparallel case,  it is
evident from Eq.\ (\ref{hp}) that $\bm{h}^{ij}_\perp=-\bm{h}^{ji}_\perp$ and
hence there are no parallel fields $\bm{h}^{ij(ji)}_{\parallel}$ able to lead
to a uniform factorizing field for the pair,  {\it unless}
$\bm{h}^{ij}_\perp=\bm{0}$ (for instance,  anti-parallel alignment directions
along the $z$ axis fulfill Eqs.\ (\ref{2v}) if $J_x^{ij}=-J_y^{ij}$ and lead to
$\bm{h}^{ij}_\perp=\bm{0}$). \qed

As previously discussed, in the uniform case $\bm{n}_i=\bm{n}_j$ the
perpendicular fields are equal and $h^{ij}_\parallel=h^{ji}_\parallel$, with
the strength of $h^{ij}_{\parallel}$ remaining {\it arbitrary}. However, when
$\bm{n}_i\neq\bm{n}_j$  Eq.\ (\ref{d1}) will lead to fixed values of the
parallel fields, which we now proceed to explicitly determine.

When $\bm{U}^{ij}$ and $\bm{V}^{ij}$ are linearly independent, by solving Eq.\
(\ref{d1}) it is found that $h^{ij}_\parallel=
-j^{ij}S[D_{i}+\bm{n}_{i}\cdot(n_{jy}n_{jz},n_{jx}n_{jz},n_{jx}n_{jy})]$. The
uniform factorizing field
$\bm{h}^{ij}_{s}=\bm{h}^{ij}_{\parallel}+\bm{h}^{ij}_{\perp}$ becomes then
($\mu,\nu,\sigma$ indicate three distinct principal axes)
\begin{equation}
({h}_{s}^{ij})_\mu=- j^{ij}S \alpha_{\mu\nu} \alpha_{\mu\sigma}\,,\;\;
\alpha_{\mu\nu}=n_{i\mu}n_{j\nu}+n_{i\nu}n_{j\mu}\,. \label{kf}
\end{equation}
On the other hand, when $\bm{V}^{ij}=\bm{0}$, Eq.\ (\ref{d1}) leads to
$h^{ij}_\parallel=-S\frac{\bm{n}_i\cdot\bm{J}^{ij}\ast\bm{n}_i+n_{i\mu}J^{ij}_\mu
+n_{j\mu}(J^{ij}_\sigma +J^{ij}_\nu)}{n_{i\mu}+n_{j\mu}}$
if $\bm{n}_{i}$ and $\bm{n}_j$ belong both to the
principal plane $\mu\nu$, and
$h^{ij}_\parallel=h^{ji}_\parallel=-S(J^{ij}_\mu+J^{ij}_\nu)$
if  $\bm{n}_{i}$ is the reflection of $\bm{n}_{j}$ with respect to the
principal plane $\mu\nu$. In this case, Eqs.\ (\ref{hnus}) leads to
\begin{equation}
({h}^{ij}_{s})_{\mu(\nu)}=-S(J_{\mu(\nu)}^{ij}+J_\sigma^{ij})n_{i\mu(\nu)}\,,
\;\;({h}^{ij}_{s})_\sigma=0\,,
\label{htpp}
\end{equation}
meaning that the $\sigma$ component of $\bm{h}^{ij(ji)}_{\parallel}$ and
$\bm{h}^{ij(ji)}_{\perp}$ cancel each other such that the local uniform
factorizing field  belongs to the principal plane $\mu\nu$.\\

{\bf  Lemma 5.} {\it Proof.} Let $\bm{h}^i$ and ${\cal J}^{ij}$ be the fields and
couplings for which the separable state $|\Theta\rangle$ (not necessarily
uniform) is a nondegenerate GS. Then, if $\bm{h}^i\rightarrow
\bm{h}^i+\delta\bm{h}^i$ and $J^{ij}_{\mu\nu}\rightarrow J^{ij}_{\mu\nu}+\delta
J^{ij}_{\mu\nu}$, the perturbed GS is $|{\rm
GS}\rangle=|\Theta\rangle+\delta|{\rm GS}\rangle$, with
\begin{eqnarray}
\delta|{\rm GS}\rangle&\approx&{\sum_\nu
{\textstyle\frac{\langle \nu|(\sum_i \delta \bm{h}^i_{\perp}\cdot\bm{S}_i+\sum_{i< j,\mu,\nu}
\delta J_{\mu\nu}^{ij}S_i^\mu S_j^\nu)|
\Theta\rangle}
{E_\nu-E_{\Theta}}}|\nu\rangle}\nonumber\\
&=&{\textstyle(\sum_i \alpha_i{S}_i^{-'}+\sum_{i,j}
\beta_{ij} {S}^{-'}_i{S}^{-'}_{j}+\ldots)|\Theta\rangle}\;\;\;\,,\label{pert}
\end{eqnarray}
up to lowest nonzero order, where $|\nu\rangle$ are the exact excited
eigenstates at the factorizing point ($H|\nu\rangle=E_\nu|\nu\rangle$, $\langle
\nu|\Theta\rangle=0$), normally entangled, $\delta\bm{h}^i_{\perp}$ is the
component of $\delta\bm{h}^i$ orthogonal to $\bm{n}_i$ and ${S}_i^{\pm'}=R_i
S_i^\pm R_i^\dagger$ the rotated spin operators ($S_i^{\pm}=S^x_i\pm \imath S_i^y$), such that
$S_i^{+'}|\Theta\rangle=0$ $\forall$ $i$. In the rotated standard basis
$\{\otimes_i|{k'}_i\rangle\}$
(${S}^{z'}_i|{k'}_i\rangle=(S_i-k)|{k'}_i\rangle$), such that
$|\Theta\rangle=|0'\rangle$,  and considering first $S_i=1/2$ $\forall$ $i$,
Eq.\ (\ref{pert}) leads, to lowest order  in the perturbations
(terms quadratic in $\alpha_i$, $\beta_{ij}$ discarded)
to a reduced pair state of the form
\begin{equation}
\rho_{ij}\approx
\begin{pmatrix}1&\alpha_i & \alpha_j & \beta_{ij} \\\bar{\alpha}_i & 0 & 0 & 0
\\\bar{\alpha}_j & 0 & 0 &0  \\ \bar{\beta}_{ij} & 0 & 0 & 0 \\ \end{pmatrix}\,.
 \label{red}
\end{equation}
The partial transpose \cite{P.96,HHH.96} of (\ref{red}) has eigenvalues $1,0$ and
$\pm|\beta|_{ij}$ up to lowest non trivial order, so that $\rho_{ij}$ will be entangled
if $\beta_{ij}\neq 0$. And the exact coefficients $\beta_{ij}$ obtained from (\ref{pert})
are, for general perturbations $\delta \bm{h}^i$ and $\delta J^{ij}_{\mu\nu}$, not strictly
zero for any pair $i,j$ linked by successive applications of the coupling in $H$, due
to the two-spin excitations present in the exact eigenstates $|\nu\rangle$. They
can, of course, be very small for distant pairs, but not strictly
zero. For higher spins $S$, $\rho_{ij}$ will be more complex but  will still contain
a first submatrix of the form (\ref{red}). Hence, it will also be entangled if
$\beta_{ij}\neq 0$, since the partial transpose of this block is the first block of the
full partial transpose $\rho_{ij}^{T_j}$ and is non positive at lowest order, preventing
the full $\rho_{ij}^{T_j}$ from being positive semidefinite.\qed

For spin $1/2$, where the entanglement of formation $E_{ij}$ is just an
increasing function of the concurrence $C_{ij}$ \cite{W.98},
Eq.\ (\ref{red}) leads at lowest  order  to $C_{ij}\approx 2|\beta_{ij}|$.
At this order, $\alpha_i$, $\alpha_j$ in (\ref{red}) do not affect $C_{ij}$ nor the
eigenvalues of the partial transpose $\rho_{ij}^{T_j}$.

\end{document}